\documentclass[manuscript, nonacm]{acmart}
\newcommand{\todo}[1]{}

\definecolor{positive}{HTML}{948E8E}
\definecolor{negative}{HTML}{948E8E} 
\newcommand{\upwardTrend}{\textcolor{positive}{\ding{115}}}
\newcommand{\downwardTrend}{\textcolor{negative}{\ding{116}}}

\let\orgautoref\autoref
\renewcommand{\autoref}
{\def\sectionautorefname{Section}%
\def\subsectionautorefname{Section}%
\def\subsubsectionautorefname{Section}%
\def\figureautorefname{Fig.}%
\def\equationautorefname{Eq.}%
\orgautoref}

\usepackage{pifont}
\newcommand{\xmark}{\ding{56}}%

\usepackage{xspace}
\newcommand{\etal}{\textit{et al.}~}
\newcommand{\eg}{\textit{e.g.,}~}
\newcommand{\ie}{\textit{i.e.,}~}

\newcommand{\one}{({\em i})\xspace}
\newcommand{\two}{({\em ii})\xspace}
\newcommand{\three}{({\em iii})\xspace}

\newcommand{\coap}{CoAP\xspace}
\newcommand{\coaps}{CoAPS\xspace}
\newcommand{\coapcoaps}{CoAP(S)\xspace}
\newcommand{\csmaca}{CSMA/CA\xspace}

\makeatletter
\newcommand{\paragraphc}[1]{\vspace*{0.03in}\noindent{\bf #1}\hspace{1ex \@minus.2ex}}
\makeatother

\usepackage{multirow}
\usepackage{siunitx}
\usepackage{graphicx}
\usepackage{adjustbox}
\usepackage[labelformat=simple]{subcaption}

\usepackage{balance}
\usepackage{wrapfig}
\usepackage{color, colortbl}

\setcopyright{acmlicensed}
\copyrightyear{2025}
\acmYear{2025}
\acmDOI{XXXXXXX.XXXXXXX}

\begin{document}

\title[Duty-Cycling is Not Enough in Constrained IoT Networking]{Duty-Cycling is Not Enough in Constrained IoT Networking: Revealing the Energy Savings of Dynamic Clock Scaling}

\author{Michel Rottleuthner}
\affiliation{%
  \institution{Hamburg University of Applied Sciences}
  \city{Hamburg}
  \country{Germany}
}
\email{michel.rottleuthner@haw-hamburg.de}

\author{Thomas C. Schmidt}
\affiliation{%
  \institution{Hamburg University of Applied Sciences}
  \city{Hamburg}
  \country{Germany}
}
\email{t.schmidt@haw-hamburg.de}

\author{Matthias W{\"a}hlisch}
\affiliation{%
  \institution{TU Dresden and Barkhausen Institut}
  \city{Dresden}
  \country{Germany}
}
\email{m.waehlisch@tu-dresden.de}

\renewcommand{\shortauthors}{Rottleuthner et al.}

\begin{abstract}
Minimizing energy consumption of low-power wireless nodes is a persistent challenge from the constrained Internet of Things (IoT).
In this paper, we start from the observation that constrained IoT devices have largely different hardware (im-)balances than full-scale machines.
We find that the performance gap between  MCU and network throughput on constrained devices enables minimal  energy delay product (EDP) for IoT networking at largely reduced clock frequencies.
We analyze the potentials by integrating dynamic voltage and frequency scaling (DVFS) into the RIOT IoT operating system and show that the DVFS reconfiguration overhead stays below the energy saved for a single, downscaled MAC operation.
Backed by these findings, we systematically investigate how DVFS further improves energy-efficiency for common networking tasks---in addition to  duty-cycling.
We measure  IoT communication scenarios between real-world systems and analyze two MAC operating modes---\csmaca  and time slotting---in combination with different CoAP transactions, payload sizes, as well as DTLS transport encryption.
Our experiments reveal energy savings between 24\%~and~52\% for  MAC operations and up to 37\% for encrypted CoAP communication.
These results shall encourage research and  system design work to integrate DVFS in future IoT devices for performing tasks at their optimal frequencies and thereby significantly extending battery lifetimes.
\end{abstract}


\maketitle

\section{Introduction} \label{sec:intro}

Energy efficiency and effective power management are essential to most device classes~\cite{oal-stiee-14}.
High-end compute servers aim for lower operational costs.
Smartphones shall provide useful standby times with permanent connectivity.
Low-end IoT devices are designed for an even tighter energy budget with multiple years of battery life.

There are essentially two approaches to power management.
Duty-cycling, \ie the temporary shutdown of system components, and resource scaling, \ie  the adaptation of system operations to current needs~\cite{sc-dpmws-01}.
Duty-cycling minimizes energy whenever parts of the system can be turned off completely but it does not allow for gradual improvements during operation.
Resource scaling improves the utilization of system components during operation and avoids wasting system resources that are underused, \eg scaling down the CPU clock speed while waiting for access to slow mass storage.

\begin{figure}
	\resizebox{0.7\columnwidth}{!}{
		\includegraphics{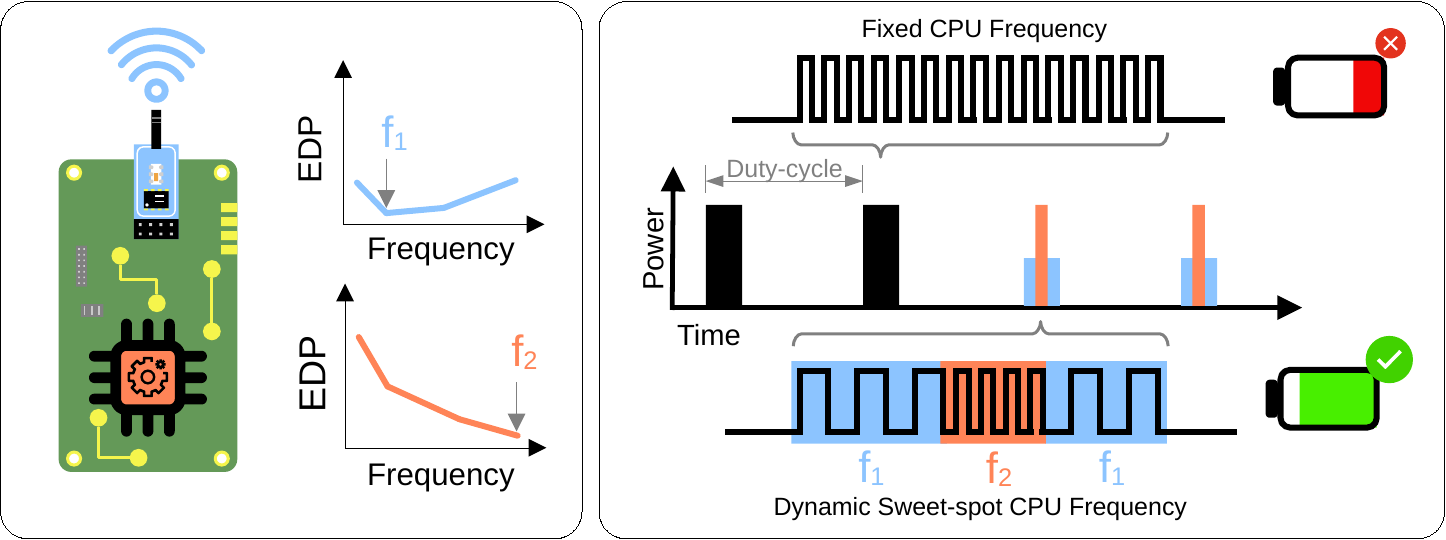}
	}
    \caption{Tasks of constrained IoT devices scale differently. Lower CPU frequencies minimize the energy and energy-delay-product (EDP) for networking tasks but high frequencies are needed to minimize the  EDP for computations. Duty-cycling alone at a fixed CPU frequency (\emph{race-to-idle}) cannot service both demands equally well. We reveal the energy savings of DVFS for widely applicable IoT networking scenarios that use duty-cycled MAC operations.}
	\label{fig:teaser}
\end{figure}

Constrained IoT devices commonly use duty-cycling to span their energy budget over the envisioned life-time.
Relying on small batteries or energy harvesting units, these devices remain mostly in deep sleep, which reduces full power consumption by about four orders of magnitude but turns off regular functionality.
Depending on the deployment use case, however, low-end embedded devices spend the majority of their energy while active.
Based on these observations, we ask how far we can minimize the consumption of online IoT devices.
We find that communicating IoT nodes can almost half their energy consumption without loss of functionality by appropriately scaling down their frequency while networking.

In this paper, we reexamine energy savings for low-end IoT devices.
We identify high potentials for Dynamic Voltage and Frequency Scaling (DVFS) during network communication due to enhanced system imbalances between MCUs and low-power wireless network interfaces (see \autoref{fig:teaser}), but also during deep sleep.
We integrate DVFS into a versatile network stack that supports a wide range of typical IoT communication scenarios.
We analyze DVFS for two energy-optimized MAC layer modes, which both do duty-cycling.
Specifically, we use Indirect Transmissions and the time-slotting Deterministic Synchronous Multichannel Extension (DSME)~\cite{IEEE-802.15.4-16} in combination with characteristic IoT networking protocols with and without encryption.
Our experiments on a commodity platform confirm energy savings of \SI{24}{\percent} for indirect transmissions, \SI{52}{\percent} for DSME, and up to \SI{37}{\percent} for encrypted \coap communication.

The current development and adoption of DVFS solutions for the IoT is in a very early stage.
Identifying the energy-optimal voltage and frequency and adapting a constrained system accordingly is challenging and faces multidimensional trade-offs.
Choosing an appropriate clock speed strongly depends on the executed task of a running application.
Measuring variable clock settings of low-power devices is subtle and affected by various hardware and software components, in particular the configured clock tree itself.
Care must be taken while identifying the proper scaling to avoid unwanted overheads or to introduce undesirable latency penalties.
Highly dynamic clock switching may in addition lead to a system behavior which is hard to predict.
We tackle these challenges.

In detail, we make the following contributions:

\begin{enumerate}
	\item We identify a performance gap in embedded systems caused by hardware imbalances and investigate the scaling behavior of characteristic  metrics for contrasting tasks. We find that energy and the energy delay product for networking minimize at reduced frequencies (\autoref{sect:problem}).
    \item We revisit MAC operations of low-power radios from an energy-centric perspective (\autoref{sec:nw-concepts}).
	\item We integrate DVFS into the RIOT OS (\autoref{sec:integration}) and measure its dynamic switching overhead. We find that the savings of \textit{a single,} downscaled MAC operation outperform the overhead (\autoref{sec:eval}).
	\item We systematically evaluate low-power wireless IoT networking with DVFS. We find significant energy savings by operating network stacks at energy-optimal frequency (\autoref{sec:eval}) and discuss the implications of our results (\autoref{sec:discussion}).
    \item We provide reproducible experiments (\autoref{sec:reproducible}) and provide in-depth empirical evidence of the energy gains in widely applicable IoT networking scenarios (\autoref{sec:eval}).
\end{enumerate}

\section{The Problem of Scaling Embedded Systems Down to Minimal Energy Consumption}
\label{sect:problem}

Scaling the system frequency does not preserve energy per se.
Potentials for saving energy depend instead on the subtle interplay between hardware characteristics and software demands.
In this section, we recap the fundamental aspects and analyze the specific properties of embedded controllers to identify sweet spots for energy optimization in the IoT.

\subsection{Fundamentals of Dynamically Reducing  Power Consumption}
Digital devices perform operations by switching the state of transistors.  
Each transition uses energy.
The two fundamental approaches to energy saving are therefore to reduce the energy per cycle, and to reduce the number of cycles.
The minimum possible energy per cycle, though is limited by the specific hardware properties of the embedded transistors, which only leaves flexibility for runtime optimizations to vary the operating parameters of voltage and frequency.
System software can optimize in two basic directions: \one operate as close as possible to the energetic (voltage, frequency) sweet-spot of the given hardware; \two reduce the number of transitions by utilizing all clock cycles to the highest capacity.

Common sources of wasteful execution are hardware-induced bottlenecks, \eg  when accessing slow IO subsystems at high CPU speed.
The efficiency is improved by aligning the processing speed to the performance of the requested subsystems.
For that, the workload of executed software must be considered, as well as the hardware-inherent imbalances between the involved subsystems.

\subsection{Characteristic Imbalances of Embedded Computing Systems}

Hardware subsystems show inherently heterogeneous performance in operation and speed, which imbalance the system in its entirety.
A well-known example is the memory-CPU speed gap of commodity computers and servers.
Embedded MCUs are less affected by this performance gap since their CPUs are much slower than high-end compute units and commonly access SRAM on the chip.

\begin{figure}[h]
    \includegraphics[width=0.65\linewidth]{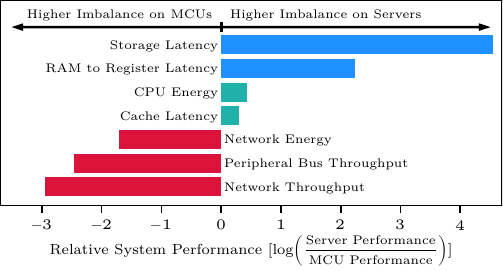}
	\caption{Comparison of system imbalances between data servers and embedded MCUs. Blue bars show pronounced server-specific performance gaps, red bars refer to the MCU performance gaps at subsystem transitions.}
	\label{fig:mcuServerComparison}
\end{figure}

In general, the degree of balance for subsystem performance varies significantly between large-scale computers and embedded devices.
\autoref{fig:mcuServerComparison} presents a detailed comparison between a modern rack server (AMD EPYC 9654) and a commodity MCU (STM32L476).
Somewhat surprisingly, the energy consumption of the CPUs (in nJ/Cycle) are very close between the server and the MCU.
In contrast, the access to peripheral buses (SPI on IoT~devices versus PCIe5 on servers) requires many more cycles per byte on an MCU than on a server.
An outstanding imbalance appears in networking (in cycles per byte), for which a server is more efficient by almost a factor of 1,000$\times$.
Correspondingly, the energy required for transmitting a byte is about 50$\times$ higher for an MCU than for a rack server.

High energy expenditure at low CPU utilization makes IoT networking ideally suited for saving energy by performance scaling.
Reducing the clock speed promises to eliminate excessive cycles and to harmonize the system balance for wireless interfaces without impairing the network performance.
As networking is a core function of embedded IoT nodes we expect that minimizing energy for communication will have a high impact on real-world deployments.

\subsection{Duty-cycling is not Enough}

Duty-cycling is the common way to enable long battery lifetime on embedded devices.
It switches off unneeded system components and can reduce the overall system consumption by more than four orders of magnitude if the MCU rests in low power mode (LPM).
During LPM the CPU clock is turned off and no instructions can be executed.
After the sleep period the CPU clock is re-enabled and execution continues at the highest possible frequency in order to return to sleep as fast as possible.
This \emph{race-to-idle} scheme is well suited for tasks that should run at full system performance, such as sensor data processing or complex control functions.
There are many scenarios, though, in which lower system performance can benefit efficiency, \eg for network transfers or slow sensor reading.
Moreover, a complex system runs various tasks which are unlikely to always match in terms of performance-to-energy scaling.
Hence, static frequency operation cannot be expected to minimize energy and the most efficient configuration is task-dependent.

An MCU that only duty-cycles between static system states will waste up to 50\% of its energy, as we will show in the following sections.

\subsection{Dynamic Voltage and Frequency Scaling}

MCUs provide a network of various system clocks on chip, which can be (re-)configured in clock trees to deliver specific clock pulses at dedicated endpoints on the chip~\cite{kbw-iruws-16,cagll-pcdmc-21,rsw-dcrci-22}.
Reconfiguring clock trees on CPUs has non-trivial impact on the energy efficiency~\cite{cbba-pcmde-10,ee-fduor-11}.
We summarize the most important aspects to cover how DVFS affects the performance and power consumption.

CPUs are CMOS based devices which consume power statically and dynamically.
Dynamic power is governed by \begin{math}P_{dyn} = \alpha \cdot C \cdot V^2 \cdot f\end{math}, with $C$ the capacitance, $V$ the voltage, $f$ the frequency, and $\alpha$ the transistor activity, \ie the probability of a transistor performing a switching operation.
Static power is given by \begin{math}P_{stat} = V \cdot I_{leak}\end{math}, where $I_{leak}$ denotes the sum of all leaking currents, such as gate leakage.
On our target MCUs the static power accounts for roughly 10\% of the total consumption during operation.
Full performance, \ie the lowest execution time, is usually obtained by running the CPU at its highest possible frequency, but lower frequencies may lead to similar performance in applications that are not able to utilize each clock cycle.
Regardless of the utilization, a higher frequency requires a higher voltage and both increase the power consumption.

Adjusting the CPU frequency has consequently two divergent effects on the energy used.
Operating at higher frequency reduces the time, during which static leakage applies, which reduces the energy spent on the static share.
But the higher voltage needed to increase the frequency will simultaneously cause higher static and dynamic consumption.
Choosing the core frequency too low will therefore waste excessive energy for static losses.
In contrast, selecting the frequency too high will waste energy due to dynamic losses without improving performance.
The sweet-spot setting finally depends on the hardware technology and the application~\cite{rd-dtgdf-17}.

\begin{figure*}[h]
	\centering
	\begin{subfigure}{0.497\textwidth}
	  \resizebox{\columnwidth}{!}{
          \includegraphics{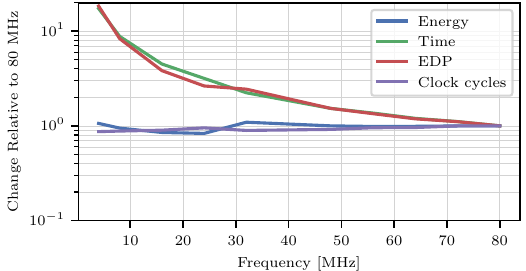}
      }
      \caption{Computing: fast Fourier transformation}
	  \label{fig:dvfs_scalability_fft}
	\end{subfigure}
	\hfill
	\begin{subfigure}{0.497\textwidth}
	  \resizebox{\columnwidth}{!}{
          \includegraphics{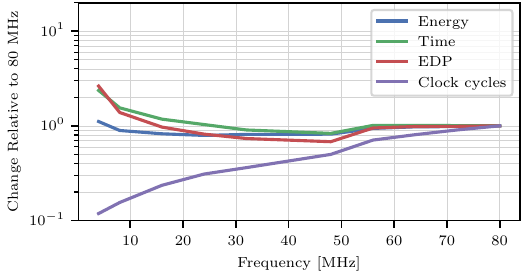}
      }
	  \caption{Networking: indirect transmission request}
	  \label{fig:dvfs_scalability_idtx}
	\end{subfigure}
	\caption{Frequency scaling of energy, execution time, energy-delay-product (EDP), and consumed clock cycles relative to their maximum frequency values for orthogonal tasks. FFT is very sensitive to  reducing CPU performance and drastically increases the energy-delay-product (EDP) at lower frequency while DVFS saves energy for network requests at low impact on the EDP.}
	\label{fig:dvfs_scalability_both}
\end{figure*}
To illustrate the difference in scalability characteristics between common computation and radio communication tasks we compare calculating a fast Fourier transform (FFT) and sending an indirect transmission (IDTX) request in \autoref{fig:dvfs_scalability_fft} and \autoref{fig:dvfs_scalability_idtx}.
The specific background on indirect transmissions and the employed target system will be detailed later in Section \ref{sec:nw-concepts} and \ref{sec:eval}, respectively.
For the FFT both the execution time and energy-delay-product (EDP) are minimal at highest frequency and significantly increase at lower frequencies, thus, energy reductions are attainable only at severe performance degradation.
The IDTX request exposes very different scalability characteristics where both EDP and energy can be reduced simultaneously by lowering the frequency.
A static frequency is therefore not able to match the sweet-spot of both operations.

Considering these fundamental properties motivates our detailed analysis in the following sections to assess how DVFS can further leverage energy-efficiency in IoT networking.

\section{Energy-efficient IoT Networking: Use Cases and Background}\label{sec:nw-concepts}

In this section, we first discuss typical networking use cases in the constrained IoT, which motivate the application scenarios we evaluate in \autoref{sec:eval}.
Next, we introduce relevant background on suitable MAC layer operations.
Our work focuses on low-power personal area networks (LoWPANs), as they are designed for energy-constrained scenarios and fully compliant with the IETF standards of a light-weight IoT network stack, \ie IPv6/6LoWPAN~\cite{RFC-4944} and UDP/\coap~\cite{RFC-7252}.

\subsection{IoT Use Cases}  
Communication strategies of IoT nodes can be classified into \one \emph{always-on}, \two \emph{normally-off}, and \three \emph{low-power}~\cite{RFC-7228}.
Our focus is on energy-constrained nodes that perform \emph{low-power} communication to \emph{appear online}.
In terms of energy management, this translates to repeatedly switching between CPU operations for processing, radio operations for frame transfers, and low-power sleep to conserve energy.
Therefore, our experiments shall cover all these system states.

To represent a wide range of applications we also consider the requirements of distinct communication scenarios.
In case of periodic sensing applications, \eg for environmental monitoring, occasional uplink-only communication is sufficient.
However, in the presence of confirmed data transmission, secure transport handshakes, or online device configurability, bidirectional communication is required.
Latency-sensitive applications additionally need reliable timing, \eg for alarm-triggering sensors or control applications with haptic feedback.
A MAC-layer which can reserve network capacity aligned to throughput and latency constraints should therefore be included.
For delay tolerant networking or firmware updates it is also relevant to incorporate data-bursts and large payload sizes which require fragmentation.

\subsection{Energy-efficient Radio Operations}
\begin{figure*}
	\resizebox{\textwidth}{!}{
		\includegraphics{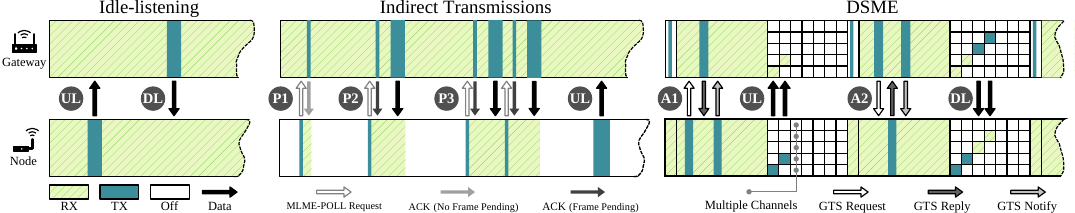}
	}
	\caption{A constrained node (bottom) communicates with the coordinator gateway (top) using different MAC-layer operating modes.
	\textbf{Idle-listening} (left), keeps both radios in RX mode to not miss incoming Up- (UL) and downlink (DL) communication and \csmaca governs channel access.
	With \textbf{Indirect Transmissions} (middle), the constrained node must explicitly request downlink frames by sending MLME-POLL MAC commands. If the ACK confirms no data pending (P1) the constrained node can turn off its radio.
	On pending data (P2), the constrained node waits in RX to receive the downlink frame. For more pending data the process is repeated (P3).
    Uplink frames are transmitted via normal \csmaca access (UL).
	\textbf{DSME} dynamically allocates guaranteed timeslots (GTS) during the CAP via \csmaca (A1) to then communicate collision free on different channels and timeslots during CFP (UL). The same process is used for downlink communication (A2, DL).
	}
	\label{fig:rxidle_vs_idtx_vs_dsme}
\end{figure*}

We now focus on IEEE~802.15.4 MAC~\cite{IEEE-802.15.4-16} operating modes which support the previously discussed use cases. \autoref{fig:rxidle_vs_idtx_vs_dsme} illustrates corresponding protocol logic and its underlying hardware operations.

\paragraphc{Idle-listening}
A radio in idle-listening mode (\ie \emph{always-on}) can operate without the CPU while waiting for data reception.
Network access happens in both directions at any time, as illustrated in \autoref{fig:rxidle_vs_idtx_vs_dsme}~(left).
Even though this mode allows the CPU to rest in low-power sleep mode until woken up by the radio, listening continuously would deplete typical batteries within a few days.
Duty-cycling the radio is therefore essential for battery life.
However, seamless communication with duty-cycled devices requires synchronization to determine when a sleepy node is ready to receive data.

\paragraphc{Indirect Transmissions}
\emph{Indirect Transmissions}, as shown in \autoref{fig:rxidle_vs_idtx_vs_dsme}~(middle), rely on downlink transmissions to be initiated by a constrained node that expects data.
Instead of continuously maintaining network time synchronization, implicit ad-hoc synchronization is obtained.
In this scheme, the coordinator (\ie an unconstrained border gateway) queues all traffic towards the sleepy node for later transmission.
The constrained node, when ready to receive data, sends a data request (\ie a short MLME.POLL MAC command packet) to the coordinator.
The gateway immediately replies with an acknowledgment frame indicating any pending data for the requesting node.
In the absence of pending data the node immediately turns the radio off.
Otherwise, it keeps its radio in listen mode to receive the pending frame. 
The coordinator sends the pending frame without further delay, keeping the listening time of the constrained node very short.
Indirect Transmissions introduce the overhead of sending data requests for each reception, and it faces similar contention issues as idle-listening in case of very busy networks.
The need for an always-on gateway permits only star topology networks but the implementation is significantly more lightweight than time slotted approaches.
\begin{figure}
	\resizebox{0.55\columnwidth}{!}{\includegraphics{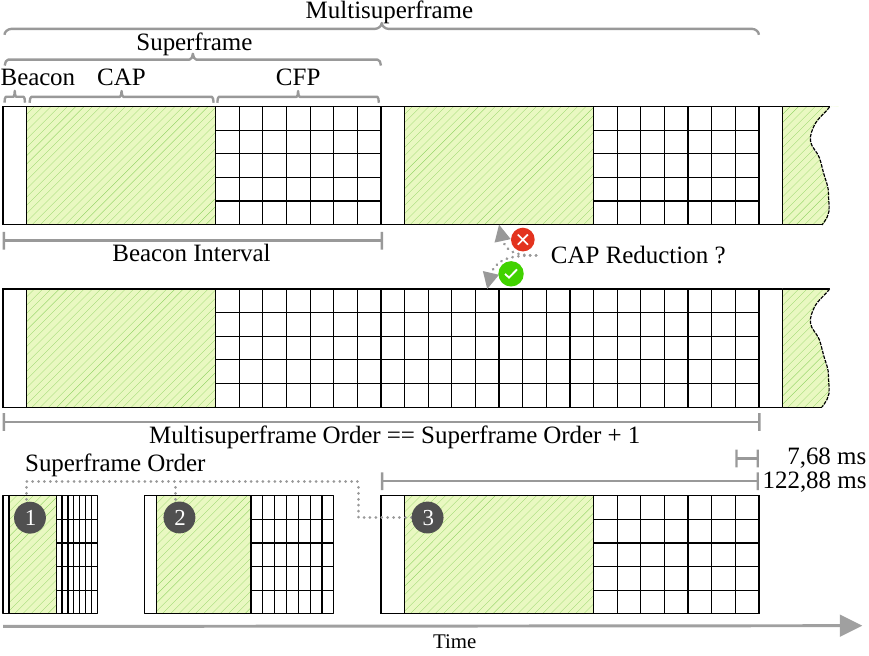}}%
	\caption{DSME adjusts to different requirements. The superframe order changes the slot duration. Multisuperframes contain a configurable number of superframes which hold a beacon slot, CAP, and CFP. The CAP can be limited to one per multisuperframe.}%
	\label{fig:dsme_slotframe_structure_overview}%
\end{figure}%

\paragraphc{Time slotted operation}%
The IEEE 802.15.4 MAC layer provides \emph{Time Slotted Channel Hopping} (TSCH) and \emph{Deterministic Synchronous Multi-channel Extension (DSME)} to transmit data in dedicated
time slots using time division multiple access (TDMA).
Both variants target a wide range of industrial and commercial applications, including reliability-, and scalability-sensitive scenarios.
They proactively synchronize the network via beacons and allow for star topologies, as well as for peer-to-peer links.
We focus on DSME as it defines slot management natively, which TSCH instead externalizes to higher layer protocols such as 6top~\cite{RFC-8480}.
DSME offers high flexibility and can be tailored  to very different application requirements.
Latency, energy, reliability, throughput, and scalability can be adjusted by selecting appropriate values for superframe order (SO), multisuperframe order (MO), and beacon order (BO).
\autoref{fig:dsme_slotframe_structure_overview} illustrates how these parameters affect the DSME superframe structure.
DSME splits communication into three distinct periods that repeat over time:
\one the beacon transmission to discover networks and maintain synchronization;
\two a contention access period (CAP) to provide \csmaca-based channel access for all nodes;
\three the contention free period (CFP), during which only guaranteed time slots (GTS) are scheduled.
GTS are negotiated between dedicated nodes and exclusively used by them.
GTS allocation can dynamically proceed during the CAP.
DSME has to activate the radio for each CAP, beacon slot, and allocated GTS. All wake-ups must be scheduled with a safety margin to bolster for clock divergence.

As features confined to the MAC, these schemes remain transparent to the upper network layers and allow for bidirectional communication.
For all schemes, we will evaluate their energetic defaults and the improvements achievable via DVFS in \autoref{sec:eval}.

\section{System Design and Implementation}
\label{sec:integration}

We now explain how the core building blocks of our approach are integrated into an IoT system.
All implementations are based on the RIOT~\cite{bghkl-rosos-18} operating system, which is a free and open source IoT operating system with a copyleft license similar to Linux.
RIOT supports about 250 different IoT boards and supplies a full-featured network stack called \emph{gnrc}.
Abstract interfaces separate each layer of \emph{gnrc} to simplify reconfigurations of the IoT protocol stack.

\subsection{IoT Network Stack Integration}

We add an implementation of the Indirect Transmission MAC feature to the \emph{netdev} abstraction of \emph{gnrc}, as by default it only supports idle-listening.
Our DSME~scenarios base on the RIOT package of openDSME~\cite{kkt-rwmnd-18}.

When evaluating the energy of a networked IoT application, it is important to consider all components of the network stack, since they are affected by DVFS system rescaling.
In our experiments, we consider the following composition of IoT networking standards.
The 6LoWPAN convergence layer enables IPv6 communication on top of IEEE 802.15.4 for all MAC modes. 
Above IPv6, we use connection-less UDP transport services and \coap as the light-weight RESTful alternative to HTTP, as well as DTLS~\cite{RFC-6347} transport layer security for authentication and encryption of \coaps.
Payloads exceeding a single frame are  \coap block-wise transferred, which splits the payload into multiple  blocked messages.
This helps to reduce adaptation layer fragmentation and increases the probability of successful message delivery~\cite{RFC-7959}.

DSME and indirect transmissions apply duty-cycling to optimize the power consumption for networking.
This increases the transmission latency, which must be considered by upper layers.
The \coap and DTLS handshake timeouts are typically set to a few seconds.
In particular, the DTLS handshake, which requires multiple round trips, is sensitive to additional delays as multiple polling intervals accumulate.
Therefore, common default values for upper layer timeouts are adjusted accordingly.
We report about specific configuration details in \autoref{sec:eval}.

\subsection{DVFS Implementation on the Operating System Level}

IoT operating systems for constrained devices commonly support low-power operation and extended battery lifetime by plain duty-cycling, only. This also holds for RIOT.
Support for dynamic clock scaling is not available out of the box and only recently attracted more attention by researchers~\cite{aasma-icdvf-20, cagll-pcdmc-21, rsw-dcrci-22, drsw-fceoc-23}.
Better abstractions and OS integration were proposed and demonstrated to work~\cite{cagll-pcdmc-21, rsw-dcrci-22} but the dynamic control is still in experimental state. This is in contrast to the widely adopted Linux implementations of CPU performance scaling~\cite{w-cps-17} and the \emph{Common Clk Framework}~\cite{t-tccf-23} hardware abstraction layer for larger devices.

Dynamic hardware clock configuration requires platform-specific low-level code and its integration into higher layer software components.
In previous work, we built an open source clock tree abstraction for online reconfiguration of constrained IoT devices~\cite{rsw-dcrci-22}.
This provides an interface for flexible switching between arbitrary clock configurations but it lacks support to efficiently integrate with short running duty-cycled operations.
The reason behind is the overhead incurred by transitioning and managing hardware-induced changes of the clock configuration.
Such hardware-induced changes occur when the system wakes up from a low power sleep mode which automatically resets the hardware clock configuration into a different setting.
The existing solution handles this transparently by always restoring the previous clock configuration.
However, if the target frequency for the next task differs, an additional frequency transition must be performed before scheduling the task.
Executing these complex transitions takes around \SI{25}{\milli\second} which is a too large overhead for short running operations, such as transmitting a single MAC command.

In this work, we implement two extensions to overcome these shortcomings: \one a mechanism to cache clock reconfiguration steps for faster execution; \two graceful handling of hardware-induced clock changes to enable faster direct transitions to the next target frequency.
Determining the current hardware state, required reconfiguration steps, and register value translations must only happen once when switching between two distinct clock configurations.
Upon the first execution a reconfiguration procedure can be translated into a list of operations that can later be executed unconditionally in sequence to repeat the same transition again.
Since each cached sequence is specific to one pair of a source and target setting, this increases memory requirements for each  frequency in use.
However, in practice a low number of frequency levels is sufficient and the number of cache entries can be limited to a compliant count using the slower execution variant as fallback.

\subsection{DVFS Control Methods}
Our experiments are intentionally designed to be widely agnostic to specific DVFS control methods. For completeness, we now discuss multiple  options and their implications.
Whether a specific control method is suitable for a particular use case strongly depends on the application requirements and runtime dynamics.

\paragraphc{Offline Profiling with Static DVFS Assignment}
Offline profiling requires extensive lab testing prior to deployment.
It allows for using measurement equipment with very high accuracy and is very light weight to implement on the target device as no runtime instrumentation is required.
The deterministic behavior makes static assignments easier to validate for hard real time guarantees~\cite{drsw-fceoc-23}.
However, static assignments lack the flexibility of adapting to runtime conditions, which limits them to applications of repetitive tasks that follow a fixed schedule.

\paragraphc{Implicit Assignment for I/O-bound Operations}
Rate limited I/O operations constitute a major bottleneck in embedded applications.
Therefore, another DVFS control method is to reconfigure clocks whenever accessing slow I/O buses~\cite{cagll-pcdmc-21}.
This is easy to implement but limited to peripheral interactions.
For very short peripheral operations the overhead of the frequency transition outweighs the energy savings.

\paragraphc{Online Scaling Analysis}
In our previous work on clock tree abstractions, we proposed a method that enables a system to self-assess the preferred CPU frequency of a task during runtime~\cite{rsw-dcrci-22}.
This concept builds on the observation that the core frequency may have limited impact on the execution time of a task, which correlates to idle clock cycles of bottlenecked task.
The OS can autonomously profile each executed thread at different frequencies, observe the impact on the execution time and select the best frequency according to its specific scalability and performance requirements.
Due to this online assessment it reflects variable runtime conditions better than a static assignment but carries the overhead of the additional profiling steps on the device.

While this approach works well for assessing the scalability of different application loads, it remains blind w.r.t. the consumption of hardware peripherals.
A task, for example, may be correctly reduced in frequency to save CPU cycles, but slowing down operations could also delay turning off the radio, which increases power consumption.
To accurately cover such side effects the solution would need to be extended by a power model of relevant system states or a measurement based self-assessment of the overall power consumption.

In this work, we focus on duty-cycled networking scenarios, in which a significant share of the consumption is defined by the communication layers that operate periodically.
In our experiments, we assign offline determined DVFS schedules statically per thread.
When the system wakes up from sleep the scheduler immediately configures the target frequency of the next runnable task instead of initializing the system to its default clock setting.
An extended online approach could potentially further minimize the energy consumption but we leave this for future work.

In the following, we analyze the energy-efficiency of our approach in different communication scenarios and measure its energetic improvements at varying clock configurations.

\section{Evaluation}\label{sec:eval}

In this section, we validate our approach with detailed power measurements while dynamically scaling system performance via DVFS.
We inspect the impact from the perspective of common network operations.
This includes different power management levels for the MCU and radio operations, \eg idle, low-power sleep, and radio listening.
We also explore clock source settings, which previously showed relevant impact in addition to the core frequency~\cite{cagll-pcdmc-21, rsw-dcrci-22, drsw-fceoc-23}.
For network operations, we separately investigate different \emph{appear online} MAC modes, constrained application methods for retrieving and sending data, and communication with and without encryption.

\subsection{Experiment Setup}
We use two devices under test (DUT) running a custom, open-source firmware that extends RIOT (see \autoref{sec:integration}).
One device operates as the always-on network coordinator.
It is permanently powered via USB and operates at the static default frequency of \SI{80}{\mega\hertz}.
The second DUT represents a constrained battery-powered device.
Both use an evaluation board from STMicroelectronics~(STM32-L476RG) and an AT86RF233 radio module.
Default settings of the RIOT device driver are used for all experiments, such as TX power and RX sensitivity, unless specified otherwise.

A static external supply voltage powers the constrained DUT with constant \SI{3.3}{\volt} using the Rhode \& Schwarz NGT20 power supply.
The positive line from the supply is serially connected to a highly accurate digital sampling multimeter (Keithley DMM7510) and its output is forwarded to the MCU input.
We apply voltage scaling via the software controlled MCU-internal voltage regulator.
For each configurations we apply the lowest voltage setting which is supported by the manufacturer specification.

For protected communication scenarios we encrypt \coap requests with the DTLS implementation of the \emph{tinydtls} RIOT package using AES and pre-shared keys.
We use a polling interval of \SI{1}{\second} for Indirect Transmissions and a slot duration of \SI{7.7}{\milli\second} for DSME.

\subsection{Dynamic Switching Overhead}
\begin{figure}%
    \includegraphics[width=0.55\linewidth]{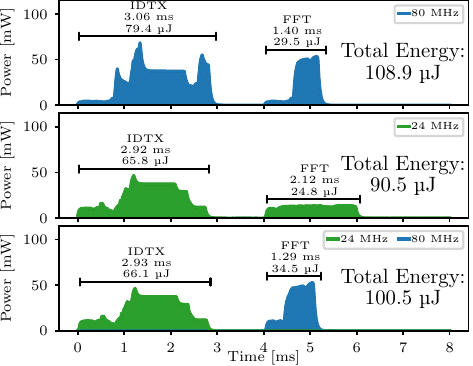}%
    \caption{Power traces for fixed and dynamic frequency operation of an IDTX request and FFT computation.}%
	\label{fig:trace_compare_zoom}%
\end{figure}

When scaling down a system, DVFS potentially extends the execution time of tasks and may add the delay of the frequency transition.  
We now quantify this overhead on our system by power trace measurements of two duty-cycled tasks that execute at different CPU frequencies.
\autoref{fig:trace_compare_zoom} compares fast and slow static frequency operation with dynamic frequency scaling.
One task performs \textit{one} indirect transmission request (left), which is the smallest MAC-layer primitive for maintaining downlink connectivity.
The other task (right) computes a fast Fourier transformation (FFT), representing a common operation of signal processing applications.
With a static frequency of \SI{80}{\mega\hertz} (top) the system consumes about \SI{20}{\percent} more energy compared to the static low frequency case with \SI{24}{\mega\hertz} (middle).
Lowering the frequency does not slow down the radio operation but the FFT execution time increases by more than \SI{50}{\percent}.
The bottom case dynamically adjusts the system to a task-specific frequency to combine the energy-efficient radio operation with the high-performance computation of the FFT.

The additional reconfiguration overhead of \SI{5}{\micro\joule} for the FFT (\SI{34.5}{\micro\joule} vs.  \SI{29.5}{\micro\joule}) remains below the energy saved from just a single MAC request  (\SI{66.1}{\micro\joule} vs.  \SI{79.4}{\micro\joule}).\footnote{Note, the phase-locked loop (PLL) initialization is faster (at higher power consumption) when the system wakes up from \SI{24}{\mega\hertz} because the most efficient PLL setting for \SI{80}{\mega\hertz} uses a slower frequency at the source oscillator.}
We note that a DVFS adaptation for a minimal network request comprises the \textit{worst case scenario} and conclude that rescaling network tasks is beneficial even then.
For longer running tasks (\eg transmission of full-size network frames or larger computations) the relative impact of the overhead will be less relevant.
These results confirm that  DVFS can be effectively utilized  in our system even for short running, duty-cycled operations as required for several networking scenarios.

\subsection{System Operation Baseline}
We now answer the question \emph{``Can DVFS reduce power consumption in low-power system states commonly selected by duty-cycling?''}.
We measure current consumption of the system in continuous receive mode (\ie idle-listening) without sending or receiving data.
We consider scenarios in which the MCU low-power mode (LPM) is turned on and off, and vary the MCU frequency for different clock sources, \ie running the core clock from the multi-speed internal RC oscillator directly or routing the clock through the PLL.
The DUT runs each configuration for \SI{10}{\second}.

\begin{figure*}[h]
	\centering
	\begin{subfigure}{0.497\textwidth}
	  \resizebox{\columnwidth}{!}{
	  \includegraphics{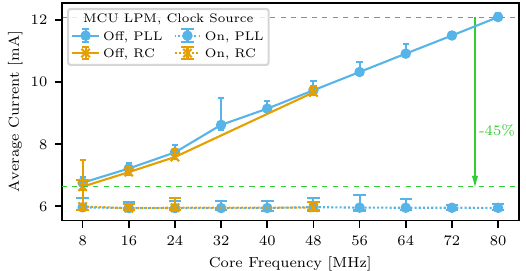}
      }
	  \caption{Radio permanently enabled (idle-listening)}
	  \label{fig:idle_listen_mculpm_on_off}
	\end{subfigure}
	\hfill
	\begin{subfigure}{0.497\textwidth}
	  \resizebox{\columnwidth}{!}{
	  \includegraphics{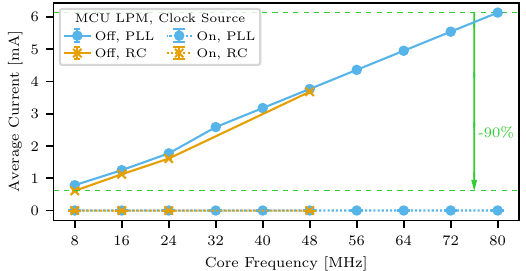}
      }
	  \caption{Radio permanently disabled}
	  \label{fig:radio_off_mculpm_on_off}
	\end{subfigure}
	\caption{Average power consumptions at different CPU frequencies, core clock sources, and MCU low power mode configurations (on/off) with the radio enabled or disabled. The RC oscillator only supports a subset of all frequencies.}
	\label{fig:fig:radio_off_on_mculpm_on_off}
\end{figure*}

Each measurement is repeated 20 times.
We present averages and variance including outliers.
\autoref{fig:fig:radio_off_on_mculpm_on_off} compares the average current consumption and observed variance when the radio is permanently enabled or disabled.
In all configurations, enabling the low-power mode yields the best performance.
Reducing the CPU frequency is clearly beneficial to approximate minimal power consumption.
It is worth noting that in contrast to the low-power sleep mode, a clocked-down IoT node is still able to slowly process data, \eg delivered via low data-rate network access.

In the idle-listening state, see \autoref{fig:idle_listen_mculpm_on_off}, all OS tasks and MCU peripherals are idle apart from mandatory overflow handling of the low-power timer.
The timer maintenance occupies negligible runtime and is required for tracking time and scheduling wake-up events.
Larger divergences from the average current appear at 8~and~\SI{32}{\mega\hertz} without LPM.
Further experiments confirmed that the presence of unrelated WiFi interference significantly contributes to the observed variances.
We expect this to be caused by power optimizations autonomously performed by the radio while awaiting packets.
The data sheet~\cite{AT86RF233-14} denotes that the effective power reduction of this optimization varies with operating conditions such as ``traffic, temperature, channel noise, and [radio] frequency settings''.
As this optimization can reduce the power consumption by over \SI{50}{\percent} and in RIOT it is enabled by default, we expect that most applications would also use this setting.

When the radio is completely disabled, as shown in \autoref{fig:radio_off_mculpm_on_off}, the IoT node does not only consume lower current, but also gains relatively more than the idle-listening configuration.
This configuration shows the isolated base consumption of the MCU at each frequency.
Without radio operation the previously seen variances disappear.
Below \SI{32}{\mega\hertz}, there is a noticeably steeper current drop because of the lower internal core voltage level supported in this region (\SI{1}{\volt} instead of \SI{1.2}{\volt}).
The clock source selection (\ie \texttt{RC} vs. \texttt{PLL}) becomes more relevant in this operating mode.

\begin{figure}%
    \includegraphics[width=0.5\linewidth]{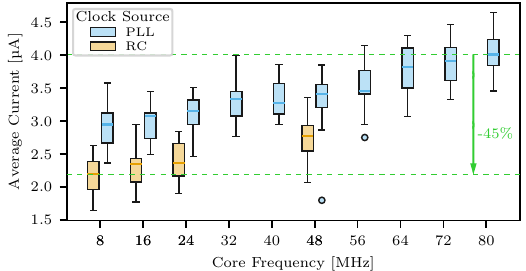}
	\caption{Zoom of \autoref{fig:radio_off_mculpm_on_off} with rescaled $y$-axis from \SI{}{\milli\ampere} to \SI{}{\micro\ampere}: Average power consumption at different core clock sources and frequencies with radio permanently disabled and MCU low power mode enabled. Boxes span Q1-Q3, median is highlighted, whiskers mark greater than Q1-1.5*IQR and less than Q3+1.5*IQR, outliers are marked with dots. The RC oscillator only supports a subset of all frequencies.}
	\label{fig:radio_off_mculpm_on}
\end{figure}%

To highlight the minimal energy requirements of the MCU, we rescale the average current from \SI{}{\milli\ampere} to \SI{}{\micro\ampere} in \autoref{fig:radio_off_mculpm_on}.
At this point, two features become clearly visible.
First, even when deep sleep duty-cycling is used and the MCU does only the bare necessities for timer overflow handling, the energy consumption can be reduced by \SI{45}{\percent}.
Second, the clock source selection becomes fundamentally important when aiming for lowest current.
For mostly asleep energy-constrained devices, this operation mode is critical as it becomes a more dominating share of the overall consumption.
The relatively rare and very short wake up events (once every second, for a fraction of a millisecond) affect the average current considerably, because of the difference in consumption between the active and low-power mode, which easily reaches more than 10,000$\times$.

Given a need for light processing or other reasons against deep sleep, it is important to reduce the CPU~frequency dynamically.
\textit{The key takeaway of this analysis is that DVFS reduces power consumption by at least 45\%, regardless of whether deep sleep duty-cycling is used.}

\subsection{Impact on Advanced MAC-layer Modes}
We now answer the question \emph{``Can we further reduce power consumption of already power-optimized MAC-Layers by using DVFS?''}.
Those MAC-layer modes increase the CPU utilization of nodes but reduce the listening time of the radio.

\paragraphc{Reducing power consumption of Indirect Transmissions}
The Indirect Transmission mode provides the most light-weight duty-cycled networking capability.
The system briefly enables the MCU and the radio once per second to poll data from the coordinator with a MAC command.
After the reply, the system immediately returns both subsystems back to sleep.
\begin{figure}%
    \includegraphics[width=0.5\linewidth]{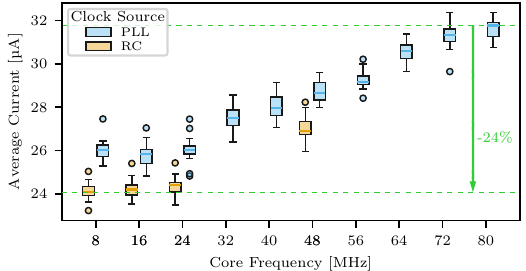}
	\caption{Power consumption at different core clock sources and frequencies when issuing MLME.POLL requests every second with MCU low power mode enabled. Boxes span Q1-Q3, median is highlighted, whiskers mark greater than Q1-1.5*IQR and less than Q3+1.5*IQR, dots mark outliers. The RC oscillator only supports a subset of all frequencies.}
	\label{fig:radio_mlmepoll_mculpm_on}
\end{figure}%
\autoref{fig:radio_mlmepoll_mculpm_on} depicts the average current consumption of Indirect Transmissions which is reduced by roughly \SI{19}{\percent} when using a lower frequency and an additional \SI{5}{\percent} when using the clock source directly instead of the PLL.

\paragraphc{Reducing power consumption of DSME}
The DSME MAC-layer is more complex than Indirect Transmissions.
Maintaining synchronization, time triggered MAC-events, and long listening periods during the contention access period results in an overall higher consumption.
To optimize for different application requirements DSME parameters can be fine-tuned for energy, reliability, and scalability.

In the following experiment, we balance our configuration between recommendations in the 802.15.4e-2012 standard and preferences for good 6LoWPAN compatibility.
For energy critical applications the standard recommends \emph{SO=1} and \emph{MO=BO=14}.
This implies a multisuperframe duration of over \SI{250}{\second} with beacons in the same period, which is very high compared to the \SI{60}{\second} 6LoWPAN reassembly timeout and default timeouts given by the \coap protocol (\texttt{MAX\_LATENCY}=\SI{100}{\second} and \texttt{MAX\_TRANSMIT\_SPAN}=\SI{45}{\second}).
Moreover, \emph{SO} must be at least 3 to accommodate full sized 802.15.4 frame transmissions within a single slot \ie a PHY service data unit (PSDU) of \SI{127}{\byte}.
We therefore set the DSME parameters to \emph{SO=3} and \emph{MO=BO=10}, which are applicable to energy constrained devices, but still allows seamless IP/\coap communication with acceptable latency.
This results in a slot duration of ~\SI{7.7}{\milli\second}, a multisuperframe duration of ~\SI{15.7}{\second}, and one beacon per multisuperframe.

\begin{figure}%
	\center
	\begin{subfigure}{0.497\textwidth}
	  \resizebox{\columnwidth}{!}{
	  \includegraphics{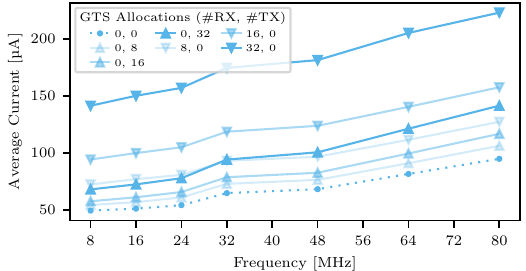}
      }
	  \caption{Absolute measures}
	  \label{fig:radio_dsme_absolute}
	\end{subfigure}
	\hfill
	\begin{subfigure}{0.497\textwidth}
	  \resizebox{\columnwidth}{!}{
	  \includegraphics{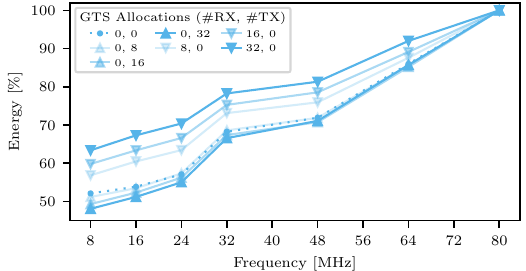}
  }
	  \caption{Relative measures}
	  \label{fig:radio_dsme_relative}
	\end{subfigure}
	\caption{Consumption of a DSME node with different numbers of guaranteed time slots allocated, running different frequencies. Relative measures are normalized to the same slot allocation group at default frequency. }
	\label{fig:radio_dsme}
\end{figure}%

\autoref{fig:radio_dsme_absolute} shows that lower core frequencies significantly affect the average current consumption, depending on the different guaranteed time slots (GTS).
Absolute energy savings maximize when the system idles, \ie no GTS are allocated (dotted line).
Allocating GTS enables communication during the contention free period (CFP) and noticeably increases the consumption with each slot.
Unused downlink slots (\downwardTrend) increase the consumption more significantly than uplink slots (\upwardTrend) because downlink transmissions need to leave the radio enabled for potentially arriving frames.
Within an uplink slot, however, a node is not required to activate the radio if no data awaits sending.
Correspondingly, the radio can be shut down before the uplink slot expires if data sending completed.

Relative energy savings differ from absolute measures in particular in uplink scenarios, see \autoref{fig:radio_dsme_relative}.
In uplink scenarios, more slots enable substantially higher energy savings because the radio can be turned off and the remaining required energy depends on the frequency.
In idle downlink scenarios, however, the required energy dominantly depends on keeping the radio awake, thus relatively less energy can be saved by reduced CPU~frequencies.

\emph{The key takeaway from this analysis is that DVFS reduces system energy even for minimal MAC operations and for all MAC layers. Savings are higher (52\%) for synchronized DSME and reduce to 24\% for asynchronous Indirect Transmissions.}

\subsection{Impact on Application Layer Scenarios}
\label{subsect:app_layer_impact}
We now answer the question \emph{``whether lower core frequencies should be used for application-layer protocols and for CPU-intensive crypto operations?''}.
We start with a basic CoAP scenario, in which encryption is disabled, and continue with CoAP over DTLS using AES encryption with a pre-shared key.
We consider both GET and POST requests.
To split bigger payloads into \SI{64}{\byte} chunks and limit 6LoWPAN fragmentation, block-wise transfer is used.
We disable retransmissions to ascertain that we always compare against the same number of transmissions.
This mostly isolates the impact of interference.
In case a transaction fails due to a collision, we simply repeat the experiment.
For DSME, 32 GTS are allocated in each direction starting with TX and RX slots in alternating order with pairs equally spaced across the CFP.
For indirect transmissions, burst of 10 consecutive \coap requests are issued.
In case of DSME, bursts are limited to four requests to guarantee they finish within the same CFP and avoid skewing results via the unrelated CAP part.
Measurements start right before the application layer instructs the \coap library to issue a request and ends when receiving the final response packet on the application layer.
Respectively, given values include all energy used by the whole system from the start of the request to its completion.
Values are normalized per request.

We emphasize that results shall be interpreted with respect to the specifics of both MAC-layer modes. 
The Indirect Transmission measurements include all MAC operations required for enabling transparent upper layer communication, including the poll request for each downlink frame.
DSME, on the other hand, performs  operations sequentially with the requests, such as beacon synchronization and the CAP.
As the ratio between CAP and CFP can be adjusted to meet diverse requirements, we isolate the \coap communication during CFP.
Therefore, results depend less on application-specific settings and are more generally applicable also to other CFP settings.

\begin{figure*}[h]
	\centering
    \begin{subfigure}{0.497\textwidth}
        \includegraphics[width=\textwidth]{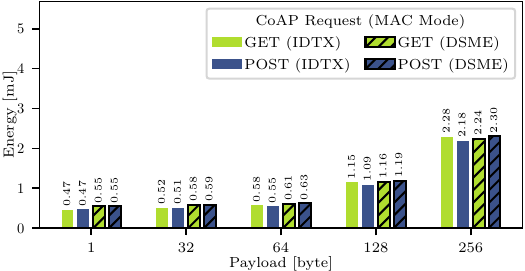}
        \caption{Unencrypted CoAP.}
        \label{fig:coap_get_post_abs_energy_idtx_mculpmon}
    \end{subfigure}
	\hfill
    \begin{subfigure}{0.497\textwidth}
        \includegraphics[width=\textwidth]{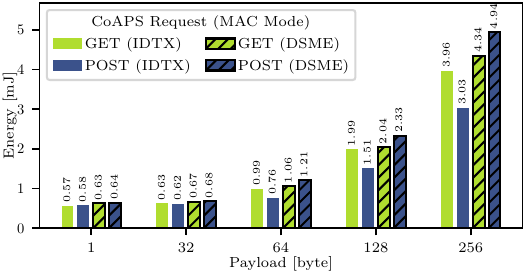}
        \caption{Encrypted CoAP (DTLS, PSK with 128 bit AES) .}
        \label{fig:coaps_get_post_abs_energy_idtx_mculpmon}
    \end{subfigure}
	\caption{Baseline energy for different CoAP(S) request scenarios at default core frequency (\SI{80}{\mega\hertz}) with MCU LPM enabled and for different payload sizes. MAC operating either in Indirect Transmission mode (\SI{1}{\second} polling interval) or DSME with low-power superframe config (SO=3, MO=10, BO=10).}
	\label{fig:coap_coaps_abs}
\end{figure*}

\paragraphc{\coap: energy requirements \underline{without} transport encryption}
\autoref{fig:coap_get_post_abs_energy_idtx_mculpmon} shows the energy consumed per unencrypted \coap request when using the default core frequency.
This measurement serves as the base reference for subsequent comparison, which will show how much of this energy is needed when running at adjusted core frequency.
DSME and Indirect Transmissions consume very similar amounts of energy for the same \coap transaction.
For smaller payloads (1-\SI{64}{\byte}), which fit into a single frame, the exact size has only subordinate impact on the energy consumed per transaction.
As soon as fragmentation is required, the energy increases more abruptly, as can be seen for 128 bytes and larger.
There are no significant differences between GET and POST requests.
With Indirect Transmissions, GET requests use slightly more energy than POST requests, whereas the opposite is observed with DSME.
This observation is in line with the roughly equal consumption during RX and TX, and an additional poll command per received frame for Indirect Transmissions.

Interestingly, the Indirect Transmissions approach enables high energy efficiency---despite the additional poll frames.
We ascribe this to the differing mechanisms for synchronizing sender and receiver.
The implicit ad hoc synchronization of Indirect Transmissions allows for very short time gaps between sending the poll command, turning the radio to RX and receiving the data.
Conversely, DSME reception slots must adhere to more pessimistic guard times to cope with clock drifts.

\paragraphc{\coaps: energy requirements \underline{with} transport encryption}
For encrypted communication, results are depicted in \autoref{fig:coaps_get_post_abs_energy_idtx_mculpmon}.
Here, differences between GET and POST are noticeably and more pronounced for 64 bytes and larger.
Additional overhead related to DTLS transport encryption overflows the first frame already for 64 bytes \coaps application payload.
Consequentially, 6LoWPAN fragmentation generates more frames for transmission.
For IDTX downlink messages this further amplifies as they require a separate poll command per frame.

\begin{figure*}
	\resizebox{\linewidth}{!}{
	\includegraphics{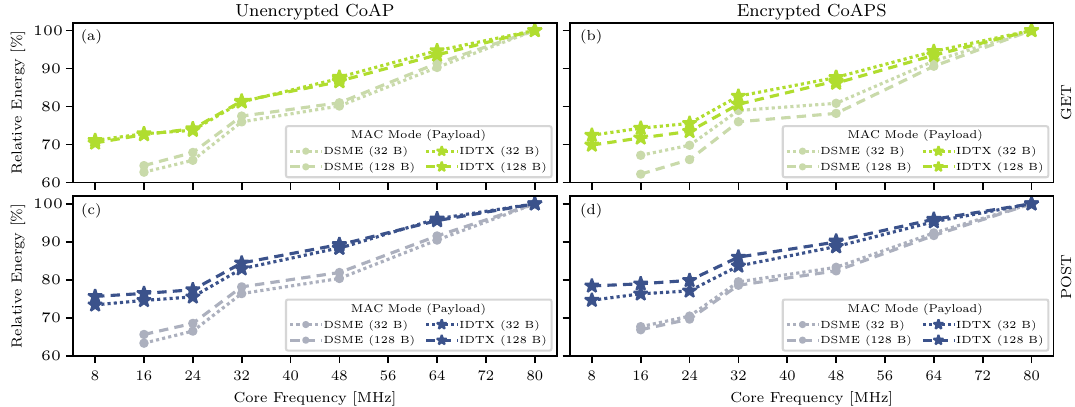}
    }
	\caption{Reduced energy levels for encrypted and unencrypted \coapcoaps GET and POST requests at lower core frequency. Using either Indirect Transmissions with \SI{1}{\second} polling interval or DSME (SO=3, MO=10, BO=10). Values are normalized to the default \SI{80}{\mega\hertz} setting of each group (see \autoref{fig:coap_coaps_abs} for reference).}
	\label{fig:idtx_dsme_coap}
\end{figure*}

\paragraphc{Reducing power consumption of \coapcoaps requests}
The previous experiments differentiated results w.r.t. fragmentation. Increasing payload sizes just add multiple CoAP blocks of similar kind to the transfer.
In the following experiment, we restrict our comparison for brevity to two related payload sizes.
\autoref{fig:idtx_dsme_coap} shows how adjusting the core frequency changes the energy spent for \coapcoaps requests.
The relative values relate to the baseline measurements presented before.
Results are normalized to the default frequency (\SI{80}{\mega\hertz}) with the same MAC operation mode and payload size.
We compare different GET and POST requests (\autoref{fig:idtx_dsme_coap}~(a) and (b) vs. (c) and (d)), encrypted and unencrypted transport (\autoref{fig:idtx_dsme_coap}~(a) and (c) vs. (b) and (d)), and Indirect Transmissions vs. DSME (markers).

Evidently, all configurations operate with higher energy efficiency at lower frequency.
For DSME, the energy reduces more significantly (35-37\%) than for Indirect Transmissions (25-30\%).
The payload size shows little effect, with encrypted variants being slightly more sensitive to changes in payload size.
The steep drop in energy at scaling down to \SI{24}{\mega\hertz} is related to the lower voltage permissible at this frequency.
For POST requests (c and d), obtainable energy savings are generally slightly lower.
Indirect Transmissions can be operated as low as \SI{8}{\mega\hertz}, whereas execution deadlines of the openDSME implementation cannot be met at \SI{8}{\mega\hertz} anymore.
We expect that optimizing the DSME implementation could lift this limitation and potentially unlock further savings.

\emph{The key takeaway from this analysis is that DVFS reduces energy consumption similarly for all CoAP and CoAPS transfers. Savings of  about 30\% for CoAP(S) GET are obtainable independent of the MAC layer.}

\section{Related Work} \label{sec:related-work}

In this section, we discuss prior work on DVFS for more powerful systems and constrained IoT devices, as well as approaches based on custom hardware in more detail.

\paragraphc{DVFS on powerful systems}
DVFS was studied from different angles, as energy requirements matter when designing computing systems~\cite{csm-ppapo-18}.
DVFS was applied to improve performance and reduce power consumption in data centers~\cite{bpopf-ieema-17}, virtual machine hosts~\cite{jhjsp-mdmpp-10}, and high-performance many-core systems~\cite{lpm-ppcpi-15, cjgpm-lpoms-16}.
Temperature-aware DVFS approaches were developed to moderate thermal throttling and reduce leakage power~\cite{kkyk-tidap-10}.
In temperature- and energy-constrained applications, DVFS was used to ensure operation within predefined limits~\cite{wqm-taesm-10}.
Similarly, heuristic algorithms were proposed to maximize performance under given power constraints~\cite{lpm-ppcpi-15}.
The impact of DVFS was analyzed for memory~\cite{shm-mprcs-12} and I/O-bound applications~\cite{mmb-ffdcm-13}
and a body of research spawned around the non-trivial problems of accurate control and performance prediction of DVFS~\cite{rlss-pppud-11, tk-pospu-23}.

DVFS performance prediction on larger systems is commonly addressed by feedback from hardware performance counters~\cite{rlss-pppud-11}, which are not available on constrained devices.
Additionally, constrained devices are very sensitive even to small overheads, which renders prior approaches for more powerful systems inapplicable.

\paragraphc{DVFS on constrained IoT devices}
Dynamic clock optimizations for constrained devices gained attention only recently~\cite{kbw-iruws-16, acilp-idvfs-17, zmxy-eicgc-17, cagll-pcdmc-21, cjkht-thepm-22, rsw-dcrci-22, drsw-fceoc-23}.
As a key insight, authors derived the need for tailoring DVFS to IoT devices, instead of purely adopting designs from powerful hardware.
For example, clocks may affect IoT time-keeping~\cite{mmsaa-taesm-12} or crypto operations~\cite{sfcf-cfiph-19} in unexpected ways.
The DVFS control on constrained devices relies on implicit clock adjustments for I/O- and compute-bound operations~\cite{cagll-pcdmc-21}, online scalability analysis~\cite{rsw-dcrci-22}, or offline schedule optimization~\cite{drsw-fceoc-23} instead of hardware-assisted performance prediction.
All of these methods could utilize a better understanding of the DVFS impact.

Up until now, the impact of DVFS on networking has not been analyzed in detail, even though networking is considered a relevant application domain for DVFS~\cite{cjkht-thepm-22}.
Prior measurements cover only selected showcases to demonstrate the validity of the proposed abstractions and control methods~\cite{cagll-pcdmc-21, rsw-dcrci-22, drsw-fceoc-23}.

In this paper, we quantify the impact of DVFS on the IoT domain by systematically analyzing its energy saving potentials in a set of widely applicable  low-power IoT networking scenarios.
Existing approaches to minimize energy by optimizing the radio transmission itself~\cite{lvs-sdwc-13, ktgs-pwblp-16, zwzzl-suira-23} do not cover the intra-system imbalances.
Our contributions are independent of specific radio technologies, which complements such existing work and allows for further prospects.
Combining ultra-low power radio with DVFS will have synergetic effects as our results show that the DVFS-agnostic radio consumption limits additional energy savings.

Chiang \etal \cite{cagll-pcdmc-21} propose a solution that optimizes I/O-bound and compute-bound tasks in terms of energy by switching between distinct slow and fast clocks.
The authors evaluate an example application that includes radio transmission without in depth analysis of the full network stack.
In our previous work~\cite{rsw-dcrci-22}, we designed a clock tree abstraction and runtime architecture for dynamic clock scaling on constrained devices.
Our evaluations focused on validating the generalized hardware abstraction, which left the thorough exploration of networking scenarios open for future work.
Dengler \etal~\cite{drsw-fceoc-23} propose an approach that optimizes the energy demand by applying offline-determined reconfiguration schedules at runtime by auto-generating static code that complies with the schedule.

\paragraphc{Custom hardware approaches}
Kulau~\etal~\cite{kbw-iruws-16} save more than \SI{40}{\percent} energy by lowering the voltage level of wireless sensor nodes below manufacturer specifications.
Mintarno~\etal~\cite{mszvc-smlep-11} tune voltage and frequency guard-bands dynamically to reduce hardware aging.
For custom chip designs, Antonio~\etal~\cite{acilp-idvfs-17} describe programmable power management for wireless sensor nodes to control DVFS and power gating in hardware.
\newcommand{\ddvfs}{D$^{2}$VFS\xspace}
For intermittent devices, a discrete variant of DVFS named \ddvfs was developed by Ahmed \etal~\cite{aasma-icdvf-20}.
It increases the number of clock cycles available for processing by reducing the frequency in reaction to the supply capacitor being drained.
Intermittent devices are orthogonal to our domain as they are too energy constrained for the standard protocol stack we consider in this work.
In contrast to custom hardware solutions, we conduct our evaluations on off-the-shelf devices and adhere to specifications.
We believe that broad hardware support and standard compliance will rather help to advance the IoT at large scale.

\autoref{relatedwork_summary_table} summarizes the closest related work and compares with our key contributions.
Accordingly, the present work covers the most comprehensive exploration of the DVFS impact on IoT networking scenarios.
\definecolor{darkgreen}{RGB}{0,190,50}
\definecolor{lightgray}{rgb}{0.9,0.9,0.9}

\newcommand{\yes}{{\color{darkgreen}\ding{52}}}
\newcommand{\no}{{\color{red}\xmark}}
\newcommand{\prt}{({\color{orange}\ding{52}})}
\newcommand{\na}{---}
\newcommand{\fna}{$^{\ast}$}
\newcommand{\fnb}{$^{\dagger}$}
\newcommand{\fnc}{$^{\ddagger}$}

\begin{table}[h]
\begin{minipage}{\linewidth}
\caption{Comparison of related contributions to networking from work most relevant to the scope of this paper in chronological order.}%
\resizebox{\linewidth}{!}{%
\begin{tabular}{lcccccccccc}
  & \multicolumn{2}{c}{Hardware compatibility} & \multicolumn{4}{c}{DVFS application} & \multicolumn{4}{c}{Networking evaluation} \\
  \cmidrule(lr){2-3}
  \cmidrule(lr){4-7}
  \cmidrule(lr){8-11}
  & COTS\fnb & MCU          & Selection \fnc & DVS & DFS & Control mechanism                                                 & MAC  & UDP & DTLS & \coap \\
 \toprule%
 Kulau~\etal~\cite{kbw-iruws-16}        & \no  & AVR (8-Bit)      & Online  & \yes & \no  & Supervising Co-processor    & \prt \fna & \no & \no & \no \\
 Ahmed \etal~\cite{aasma-icdvf-20}      & \no  & MSP430 (16-Bit)  & Offline & \yes & \yes & Frequency follows voltage   & \no   & \no  & \no & \no \\
 Chiang \etal \cite{cagll-pcdmc-21}     & \yes & ARM (32-Bit)     & Offline & \no  & \yes & Switch for I/O operations      & \prt \fna & \no  & \no & \no \\
 Rottleuthner \etal~\cite{rsw-dcrci-22} & \yes & ARM (32-Bit)     & Online  & \yes & \yes & Switch per thread              & \prt \fna & \yes & \no & \no \\
 Dengler \etal~\cite{drsw-fceoc-23}     & \yes & RISCV (32-Bit)   & Offline & \no  & \yes & Static schedule              & \no  & \no  & \no & \no \\
\rowcolor{lightgray}
 This work                              & \yes & ARM (32-Bit)     & Offline & \yes & \yes & Switch per thread              & \yes & \yes & \yes & \yes \\
 \bottomrule%
 \vspace{-0.2cm} \\
 \multicolumn{9}{l}{\fnb The Solution works on common off-the-shelf hardware without additional components.} \\
 \multicolumn{9}{l}{\fna IEEE~802.15.4 frame transmissions without covering MAC operating modes and parameters.} \\
 \multicolumn{9}{l}{\fnc The frequency selection step can happen online or offline.} \\
\end{tabular}
}%
\label{relatedwork_summary_table}%
\end{minipage}
\end{table}

\section{Discussion}
\label{sec:discussion}

\paragraphc{Why is DVFS not widely adopted in the constrained IoT?}
The key benefit of duty-cycling over DVFS is the respectable performance at deceptive simplicity.
DVFS is more complex for mainly two reasons.
First, it introduces additional software, which requires memory.
Second, it is complicated to control as there are many variables involved.
Accurately predicting their influence is non-trivial and the impact on the execution speed requires careful handling to prevent adverse effects.
On powerful systems the overhead is less concerning and the effects are well-understood, but these challenges are still open on constrained embedded devices.

\paragraphc{Does DVFS introduce notable latency penalties on IoT networking?}
No.
Even though a lower clock frequency reduces the instruction throughput, the effect on the communication latency is negligible for two reasons.
First, the evaluated energy-optimized networking operations are not CPU-limited.
Second, they are inherently time-based.
Indirect transmissions and DSME cope well with slower processing and by design (see \autoref{sec:nw-concepts}), operate with communication delays that are orders of magnitude larger than processing delays due to reduced CPU speed.
\begin{table}[h]
\caption{Maximum timing delta between different core frequencies for bursts of application layer requests (10 for IDTX, 4 for DSME).}%
\resizebox{0.5\linewidth}{!}{%
    \begin{tabular}{ lrrrr }
    \toprule
    \multirow{2}{4em}{} & \multicolumn{2}{c}{CoAP} & \multicolumn{2}{c}{CoAPS}
    \\
    \cmidrule(lr){2-3}
    \cmidrule(lr){4-5}
    time [ms]                  & tburst$_{\text{min}}$ & max $\Delta$t &  tburst$_{\text{min}}$ & max $\Delta$t\\
    \midrule
    DSME  &      2,237.0          &  1.3 & 2,238.0 &  2.2 \\
    IDTX  &      10,007.0        &  3.4 & 10,007.0  & 2.7 \\
    \bottomrule%
    \end{tabular}}%
\label{timingpenalties}%
\end{table}
A lower frequency effectively leverages slack time during less timing-critical preparation steps of a transmission and thereby prevents wasting unused CPU cycles.
However, our experiments also point out limits.
For DSME it is critical to finish preprocessing of a frame before the communication slot starts.
This requirement is not met by our test system when scaling down to \SI{8}{\mega\hertz} (see \autoref{subsect:app_layer_impact}).
Communication happens strictly within the fixed time slots otherwise and therefore the impact on communication delays is negligible.
The largest observed timing variance including the application layer was \SI{3.4}{\milli\second}, which is comparable to CSMA/CA backoff delays.
\autoref{timingpenalties} compares worst cases, \ie the minimal burst durations (tburst$_{\text{min}}$) to the maximum time penalty (max $\Delta$t), which confirms insignificance.

\paragraphc{Do results hold for other technologies than IEEE 802.15.4?}
Yes.
There are three different types of radio systems that can benefit from DVFS.
First, radio subsystems that are interfaced through bandwidth-limited serial connections, such as modems for LTE-Cat-M1 or Narrowband-IoT.
Second, network technologies that exhibit even lower data rates, such as LoRa, Sigfox, and other sub-gigahertz radios.
In both cases, the imbalance between network access and CPU are similar or even higher than in IEEE~802.15.4 and can thus be counterbalanced by DVFS.
The third class are wireless technologies that provide higher bandwidth, such as Bluetooth and ultra-wideband.
Faster networking, when considered in isolation, could reduce the CPU-to-network imbalance.
Energy requirements per byte, however, are lower with Bluetooth compared to IEEE~802.15.4.
Our results show that less energy per byte enables larger potentials for saving energy with DVFS.

\paragraphc{Do results generalize to other software and hardware?}
Yes, we anticipate qualitatively comparable results for other targets.
Even though exact results depend on hardware-software combinations, our findings provide substantial evidence that the exposed optimizations are systematically accessible.
We use clock abstractions that were multi-platform validated~\cite{rsw-dcrci-22} to assess the potential of dynamic voltage, frequency, and clock source adaptations.

Our implementation is based on the open-source RIOT operating system and uses a protocol stack of open standards, which improves applicability to a wider application domain.
Instead of investigating a single highly optimized application for a specific use-case, we incorporate a multitude of reusable software implementations from a community of different developers.
The device we use in our evaluation is based on a common ARM Cortex-M4, which is widely adopted by many manufacturers for different hardware platforms.
A benchmarking study determined our specific MCU model (STM32-L476RG) to use very low energy per clock cycle compared to other 32-bit microcontrollers~\cite{gkcak-bbiee-20}.
Systems with more wasteful clock cycles may therefore obtain even more drastic energy savings with DVFS.
Related work demonstrates comparable scaling behavior for other target devices~\cite{aasma-icdvf-20, cagll-pcdmc-21, rsw-dcrci-22, drsw-fceoc-23}, but none of them systematically analyzed and quantified the impact on IoT networking scenarios.

\section{Conclusions and Outlook}
\label{sec:c+o}

Minimizing energy consumption is of focal interest for the constrained IoT, as energy is the limiting resource in long-term deployments.
In this paper, we addressed core system properties to identify energy saving potentials.
Our work started from the observation that large amounts of energy are wasted whenever imbalanced system components interact.
We identified wireless networking as a dominant source of imbalance in common IoT use cases.

We contributed system rescaling of low-power networking as a key measure to optimize energy consumption in the IoT.
We integrated dynamic voltage and frequency scaling (DVFS) into RIOT OS and could show that the overhead of dynamic switching remains well below MAC layer savings.
We applied our approach to  a widely applicable set of low-power communication scenarios, as well as to sleep modes attained in duty-cycling.

Our findings indicate that combining duty-cycling with DVFS unlocks unprecedented energy savings of up to \SI{50}{\percent}, which are well beyond reach of pure duty-cycling.
It is particularly noteworthy that these savings apply to a system while networking and while asleep.
Our measurements also revealed energy savings of up to 37\% for (cpu-intensive) encrypted CoAP communication.

This work has shown how hidden energy potentials can be unlocked based on a systematic analysis and an adaptive system configuration.
We envision three future directions for this research. First, in-depths experiments with further IoT communication technologies should quantify their saving potentials from resource scaling.
Second, an exhaustive trace analysis of all system tasks should identify further system imbalances  and derive a comprehensive scaling policy thereof. Finally, long-term deployments with continuous resource measurements should grant insights into the gains that are effectively achievable in the wild.

\paragraphc{Artifacts}
\label{sec:reproducible}
We will make our artifacts publicly available.

\bibliographystyle{ACM-Reference-Format}
\bibliography{own,iot,energy,rfcs,layer2,internet}
\balance
\label{lastpage}

\end{document}